\RequirePackage[left]{lineno}
\documentclass[aps,prl,twocolumn,superscriptaddress,showpacs,preprintnumbers,amsmath,amssymb,letter]{revtex4-2}
\usepackage[utf8]{inputenc}
\usepackage{graphicx} 
\usepackage{dcolumn}  
\usepackage{lineno, blindtext}
\usepackage{color}
\usepackage[normalem]{ulem} 
\usepackage{comment}
\usepackage[hypertexnames=false,colorlinks,linkcolor={red},citecolor={red},urlcolor={blue}]{hyperref}
\usepackage{gensymb}
\usepackage{orcidlink}

\begin{document}

\preprint{\vbox{ \hbox{   }
}}

\title{ \quad\\[1.0cm] Search for the decay {\boldmath$B_s^0\rightarrow\pi^0\pi^0$} at Belle}
\noaffiliation
  \author{J.~Borah\,\orcidlink{0000-0003-2990-1913}} 
  \author{B.~Bhuyan\,\orcidlink{0000-0001-6254-3594}} 
  \author{I.~Adachi\,\orcidlink{0000-0003-2287-0173}} 
  \author{H.~Aihara\,\orcidlink{0000-0002-1907-5964}} 
  \author{D.~M.~Asner\,\orcidlink{0000-0002-1586-5790}} 
  \author{V.~Aulchenko\,\orcidlink{0000-0002-5394-4406}} 
  \author{T.~Aushev\,\orcidlink{0000-0002-6347-7055}} 
  \author{R.~Ayad\,\orcidlink{0000-0003-3466-9290}} 
  \author{V.~Babu\,\orcidlink{0000-0003-0419-6912}} 
  \author{S.~Bahinipati\,\orcidlink{0000-0002-3744-5332}} 
% ...
\collaboration{The Belle Collaboration}

\begin{abstract}
We report the results of the first search for the decay $B_s^0\rightarrow\pi^0\pi^0$ using $121.4\ \rm fb^{-1}$ of data collected at the $\Upsilon(5\rm S)$ resonance with the Belle detector at the KEKB asymmetric-energy $e^+e^-$ collider. We observe no signal and set a 90\% confidence level upper limit of  $7.7\times 10^{-6}$ on the $B_s^0\rightarrow\pi^0\pi^0$ decay branching fraction.
\end{abstract}

\maketitle

\tighten

{\renewcommand{\thefootnote}{\fnsymbol{footnote}}}
\setcounter{footnote}{0}

The study of heavy-flavored hadrons decaying to hadronic final states provides an important input for understanding the interplay between strong and weak interactions. These type of decays involving weak annihilation amplitudes can be a promising place to look for disagreement between theoretical predictions and experimental observations. These decays are highly suppressed and often neglected in theoretical calculations. However, the inclusion of rescattering effects into the theoretical framework naturally enhances their contribution \cite{gronau}. Recently it was observed that the predicted branching fraction for the decay $B^0_s \to \pi^+ \pi^-$, which involves topological annihilation diagrams, was substantially smaller than its measured value by the LHCb experiment \cite{LHCb_PA}. This discrepancy between theoretical prediction and experimental measurement may require some models of strong interaction processes to be revisited \cite{fit_QCDF}. In these aspects, searches for decays involving weak annihilation amplitudes become important and necessary.

\begin{figure}[htb]
\centering
\includegraphics[height=5.0cm]{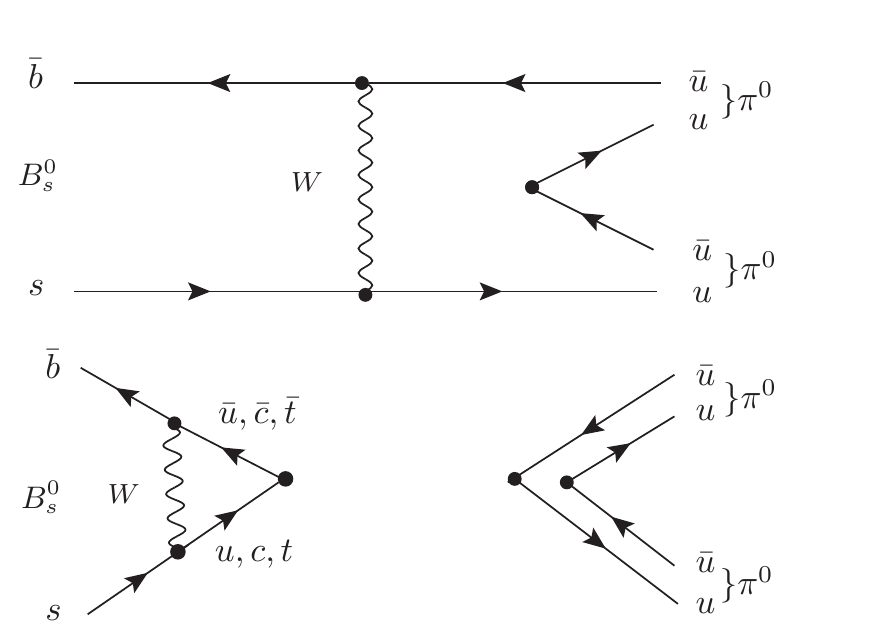}
\caption{$W$-exchange (top) and ``penguin'' annihilation (bottom) Feynman diagrams for $B_s^0\rightarrow\pi^0\pi^0$. }
\label{Topological_diagrams}
\end{figure}

Within the standard model (SM), the decay $B^0_s \rightarrow \pi^0 \pi^0$ proceeds via the $W$-exchange and ``penguin'' annihilation amplitudes, as shown in Fig.~\ref{Topological_diagrams}. Theoretical calculations based on the Flavor Diagram Approach (FDA) \cite{b2pp}, perturbative Quantum Chromodynamics (pQCD) \cite{pqcd} , and QCD factorization \cite{qcdf} predict the branching fraction ($\mathcal{B}$) to be $(0.40 \pm 0.27)\times 10^{-6}$ , $(0.28 \pm 0.09)\times 10^{-6}$, and $(0.13 \pm 0.05)\times 10^{-6}$, respectively. The only measurement for this decay was made by the L3 experiment in 1995, which reported an upper limit (UL) of $\mathcal{B} < 2.1 \times 10^{-4}$ at $90\%$ confidence level (CL) \cite{L3}. The search for the decay  $B^0_s \rightarrow \pi^0 \pi^0$ \cite{charge_conjugate} described in this Letter is based on a data sample of $121.4\ \rm fb^{-1}$ collected at the $\Upsilon(5\rm S)$ resonance using the Belle detector.

The Belle detector at the KEKB \cite{KEKB} asymmetric-energy $e^+e^-$ collider is a large-solid-angle magnetic spectrometer that consists of a silicon vertex detector, a 50-layer central drift chamber, an array of aerogel threshold Cherenkov counters, a barrel-like arrangement of time-of-flight scintillation counters, and a CsI(Tl) crystal-based electromagnetic calorimeter (ECL) located inside a super-conducting solenoid coil that provides a $1.5$ T magnetic field. An iron flux-return outside the coil is instrumented to detect $K^{0}_{L}$ mesons and identify muons. A detailed description of the Belle detector can be found elsewhere  \cite{detector1, detector2}. The analysis relies on the ECL component of the detector for the reconstruction of the photons in the $B_s^0\rightarrow\pi^0\pi^0$ decay final state.

The production cross-section of the $\Upsilon(5\rm S)$ resonance at the $e^+e^-$ centre of mass (c.m.) energy of $10.86$ GeV is $\sigma^{\Upsilon(5\rm S)}_{b\bar{b}} = (0.340 \pm 0.016$) nb \cite{Sevda}, and the fraction of $b\bar{b}$ events giving rise to $B^0_s$ production modes, $B^{(*)0}_s\bar{B}^{(*)0}_s$, is measured to be $f_s = (0.201 \pm 0.031)$ \cite{pdg2022}. There are three kinematically allowed modes of production of $B^0_s$ mesons: $B^{*0}_s\bar{B}^{*0}_s$, $B^0_s \bar{B}^{*0}_s$ or $B^{*0}_s\bar{B}^0_s$, and $B^0_s\bar{B}^0_s$. The production fractions from the former two are $(87.0 \pm 1.7)\%$  and $(7.3 \pm 1.4)\%$, respectively \cite{Sevda}, while the remaining fraction is from the $B^0_s\bar{B}^0_s$ mode. The $B^{*0}_s$ decays to $B^0_s$ by radiating a low-energy photon that is usually not identified due to its poor reconstruction efficiency. The number of events with $B^0_s\bar{B}^0_s$ is, therefore, estimated to be $N_{B^0_s\bar{B}^0_s} = 121.4 \ \rm fb^{-1}\ \cdot \sigma^{\Upsilon(5\rm S)}_{b\bar{b}}\ \cdot $ $f_s = (8.30 \pm 1.34) \times 10^{6}$.

We employ a ``blind" analysis procedure to leave out the experimenter's biases and develop our analysis strategy with Monte-Carlo (MC) samples. In a ``blind" analysis, the signal region is kept hidden until the selection criteria are finalized. The signal MC samples are generated with EvtGen \cite{evtgen} and simulated with GEANT3 \cite{geant3} to model all possible detector effects. Background studies are performed with MC samples six times larger than the integrated luminosity of data. The analysis procedure is validated with a control sample of $B_{d}^{0} \rightarrow \pi^0\pi^0$ decays produced at the $\Upsilon(4\rm S)$ resonance, which closely resembles the signal. 

We reconstruct  $B_{s}^{0}\to\pi^{0}\pi^{0}$ with $\pi^0 \rightarrow \gamma \gamma$. Photon candidates are reconstructed from ECL clusters that do not match any charged track and have energy greater than 50 (100) MeV in the ECL's barrel (end-caps) region. The forward end-cap, barrel, and backward end-cap regions of the ECL are given by $12\degree < \theta < 31.4\degree$, $32.2\degree < \theta < 128.7\degree$, and $131.5\degree < \theta < 157.2\degree$, respectively, where $\theta$ is the polar angle in the laboratory frame with respect to the detector axis, in the direction opposite to the $e^+$ beam. To remove the off-time (radiative) Bhabha and $e^+e^- \to \gamma \gamma$ events, a timing criterion based on the beam collision time is applied, which is determined at the trigger level for each candidate event. The invariant mass of the two-photon combination must lie in the range of $118 \ \text{MeV}/c^{2} < m(\gamma \gamma) < 152 \ \text{MeV}/c^{2}$, corresponding to $\pm 2.4$ standard deviations ($\sigma$) of the invariant mass resolution around the nominal $\pi^{0}$ mass \cite{pdg2022}. A mass-constrained fit is subsequently performed to improve the $\pi^0$ momentum resolution.
 
To further select the $B^0_s$ candidates, we apply selection criteria on their beam-energy-constrained mass $M_{\rm bc}=\sqrt{{(E_{\rm beam})^{2}-|\vec{p}_{\rm reco}|^{2}c^2}}/c^2$ and the energy difference $\Delta E^{\prime}=E_{\rm reco}-E_{\rm beam} + M_{\rm bc}c^2 - m_{B^0_s}c^2$ in the $e^+e^-$ c.m. frame, where $E_{\rm beam}$ is the beam energy, $\vec{p}_{\rm reco}$ and $E_{\rm reco}$ are the momentum and energy, respectively, of the reconstructed $B_s^0$ candidate. The world average value is used for the mass of the $B^0_s$ meson, $m_{B^0_s}$ \cite{pdg2022}. A $B^0_s$ candidate is retained for further analysis only if it satisfies the requirement that $5.300$ GeV/${\it c}^2$ $<M_{\rm bc}<5.434$ GeV/${\it c}^2$ and $-0.60 \ \text{GeV} < \Delta E^{\prime} < 0.15 \ \text{GeV}$.

The backgrounds near the $\Upsilon(5\rm S)$ resonance which can affect the analysis are: continuum ($e^+e^- \rightarrow q\bar{q},\  q =$  $u$,$\ d$,$\ s$,$\ c$), $B_s^{(*)}\bar B_s^{(*)}$ decays (referred as \textit{bsbs}) and $B^*\bar{B}^*$, $B^*\bar{B}$, $B\bar{B}$, $B^{*}\bar B^{*}\pi$, $B^*\bar{B}\pi$, $B\bar{B}\pi$ and $B\bar{B}\pi\pi$ ($B=B^0,B^{+}$) decays (referred as \textit{non-bsbs}). Additional background MC studies on the peaking background of the types $B^0_s \to \rho^+ \rho^-$ and $B^0_s \to K^0_s\pi^0$ show that their contributions are negligible. We also find no \textit{bsbs} and \textit{non-bsbs} background after applying all of the aforementioned selection criteria. Background MC studies, therefore, reveal the dominance of continuum background over the other types of background. Their suppression requires topological variables, which classify the signal and the continuum background based on their event shape variables in the $e^+e^-$ c.m. frame. 

In signal events, $B^0_s$ pairs are produced with small momenta, and the distribution of their decay products tends to be spherical. In contrast, the quark pairs of the continuum background are produced with a significant amount of momentum; therefore, their decay product distribution has a jet-like topology. A neural network algorithm (NN) \cite{nn} is employed to suppress the continuum background. The input of the NN includes sixteen modified Fox-Wolfram moments \cite{ksfw}, and $\cos\theta_T$ (see Section $9.3$ in \cite{bevan}) to provide additional discrimination between the signal and the continuum background. The angle $\theta_T$ is defined as the angle between the thrust axis of the signal $B^0_s$ candidate and the thrust axis of the remainder of the events. The NN was trained on MC samples with consistency checks to ensure no over-training.

The choice of the selection criterion on the output of the NN, $\mathcal{C}_{\mathrm{NN}}$, is determined based on a Punzi's figure-of-merit (FOM) optimization \cite{punzi}, where the significance level is set to three standard deviations. The $\mathcal{C}_{\mathrm{NN}}$ distributions for the continuum background and the signal lies in the range of $[-1,+1]$, where the continuum backgrounds peak at $-1$ and the signal candidates at $+1$. We require $\mathcal{C}_{\mathrm{NN}}$ to be greater than $0.90$ for this analysis. This condition removes $99\%$ of the continuum background with a signal loss of $53\%$. To facilitate the data modelling, $\mathcal{C}_{\mathrm{NN}}$ was transformed to another variable, $\mathcal{C}^{\prime}_{\mathrm{NN}}$ using the following formula
\begin{linenomath}
\begin{equation}
\cal{C}'_{\mathrm{NN}} = \log \bigg(\frac{\cal{C}_{\mathrm{NN}} - \cal{C}_{\mathrm{NN(min)}}}{\cal{C}_{\mathrm{NN(max)}} - \cal{C}_{\mathrm{NN}}}\bigg),
\end{equation}
\end{linenomath}
where $\cal{C}_{\mathrm{NN(min)}} =$ $0.90$ and $\cal{C}_{\mathrm{NN(max)}}$ is the maximum value of $\cal{C}_{\mathrm{NN}}$ obtained from the NN distribution. 

After applying the selection criteria described above, 10.3\% of
signal MC events have more than one candidate. We select the best $B^0_s$ candidate by choosing the one with the smallest sum of the $\chi^2$ of the mass-constrained fits to the two $\pi^0$s. This method chooses the correct $B_{s}^{0}$ candidate 56\% of the time. The misreconstructed fraction of events after applying all the selection criteria is found to be negligible; hence, they are not treated separately. The overall signal reconstruction efficiency in this analysis is $(12.69 \pm 0.05)\%$.

\begin{figure*}
\centering
\includegraphics[height=4.7cm]{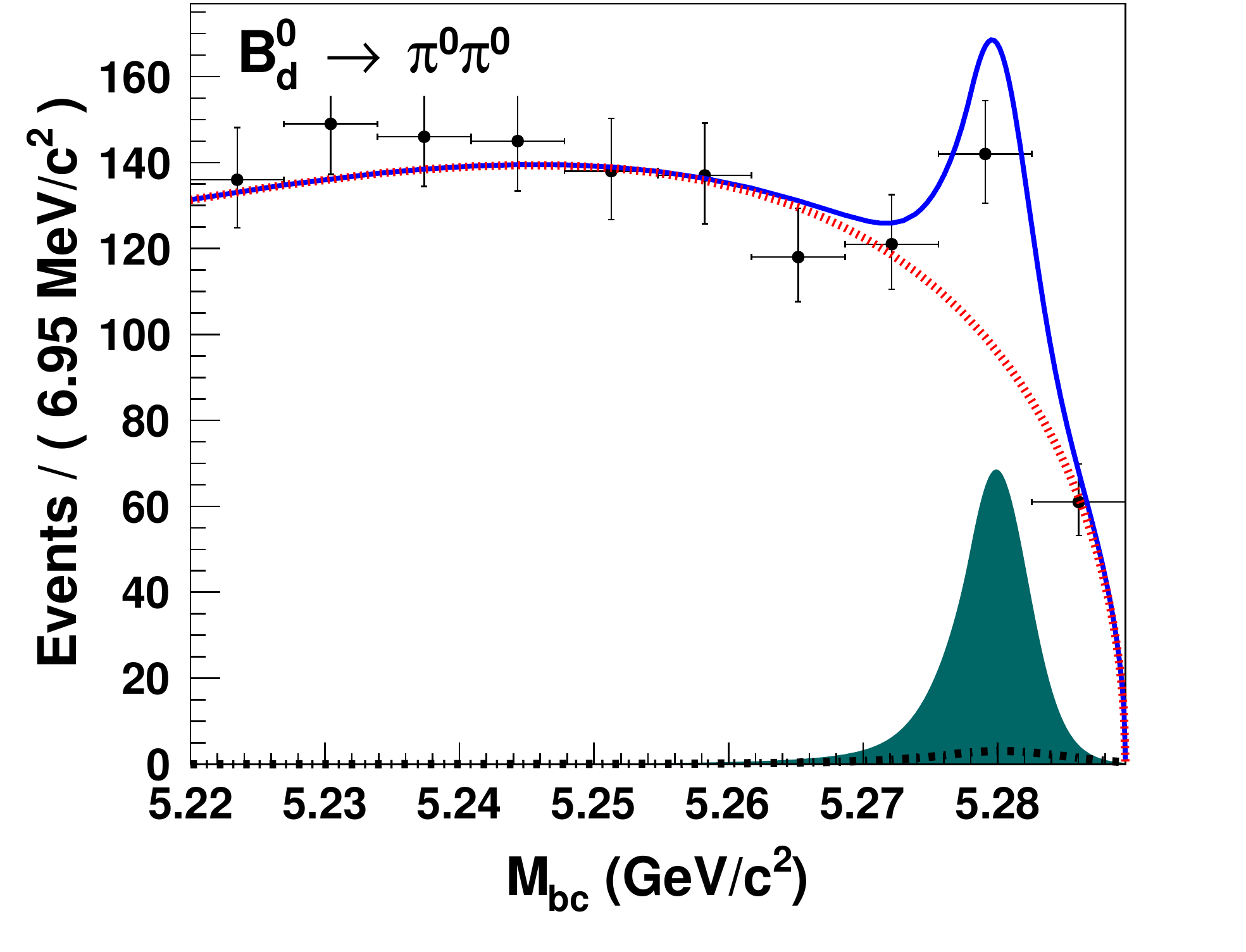}%
\includegraphics[height=4.7cm]{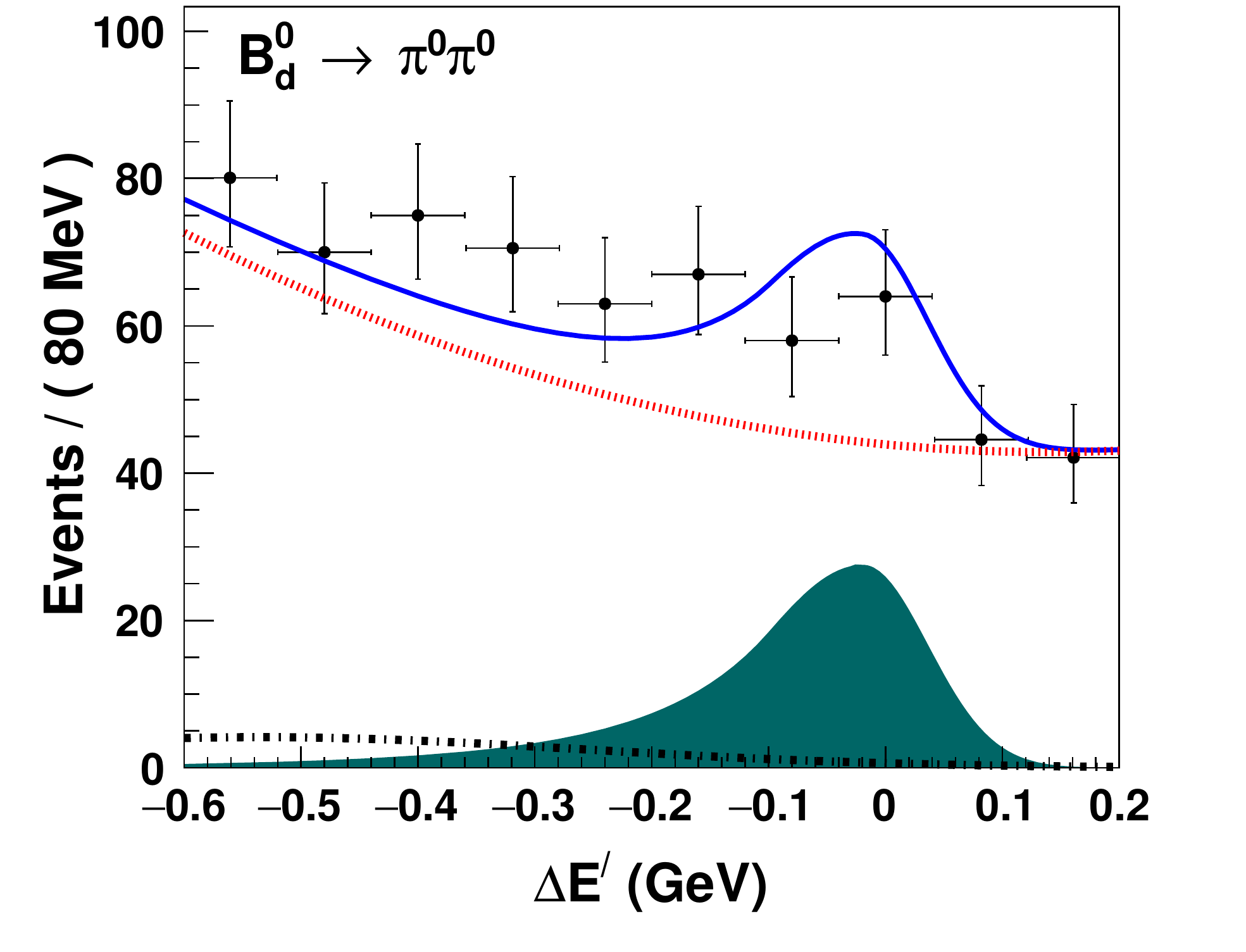}%
\includegraphics[height=4.7cm]{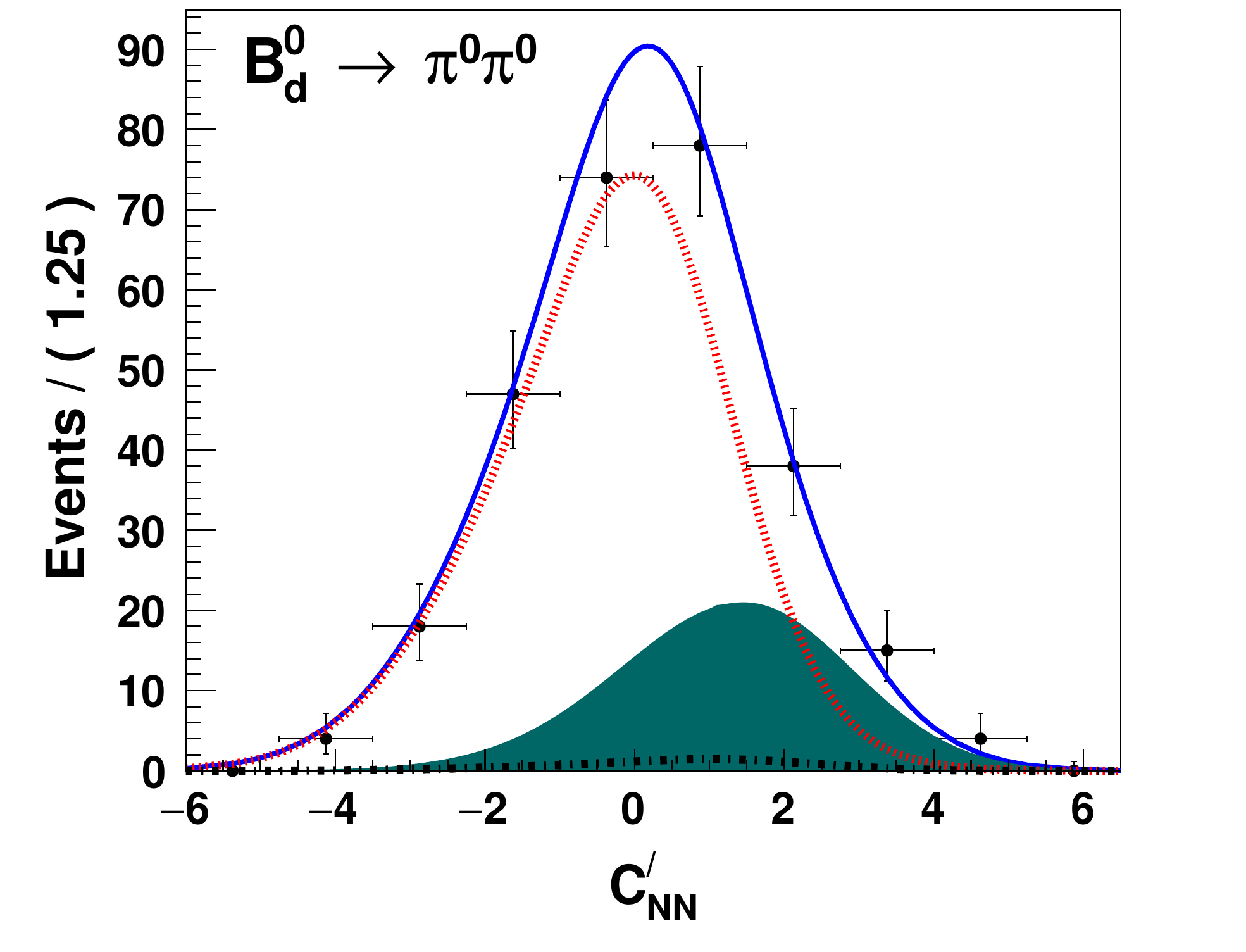} 
\caption{Signal enhanced projections of $M_{\rm bc}$ (left), $\Delta E^{\prime}$ (middle), and $\mathcal{C}^{\prime}_{\mathrm{NN}}$ (right) for the control sample, $B^0_{d} \rightarrow \pi^0 \pi^0$. Each plot is generated by applying the signal region selection criteria on the two variables other than the plotted variable. The signal regions for the three variables are as follows, $5.2700 \ \text{GeV}/c^2 < M_{\rm bc} < 5.2895 \ \text{GeV}/c^2$, $-0.23 \ \text{GeV} < \Delta E^{\prime} < 0.15 \ \text{GeV}$, and $-3.10 < \mathcal{C}^{\prime}_{\mathrm{NN}} < 7.61$. The dark-filled, red (dotted), black (dash-dotted), and blue (solid) color distributions represent the signal, continuum background, rare $B^0_d$ background (backgrounds arising due to $b \to u$ transitions) and total fit function, respectively. Points with error bars represent data.}
\label{Bdpi0pi0}
\end{figure*}

\begin{figure*}
\centering
\includegraphics[height=4.7cm]{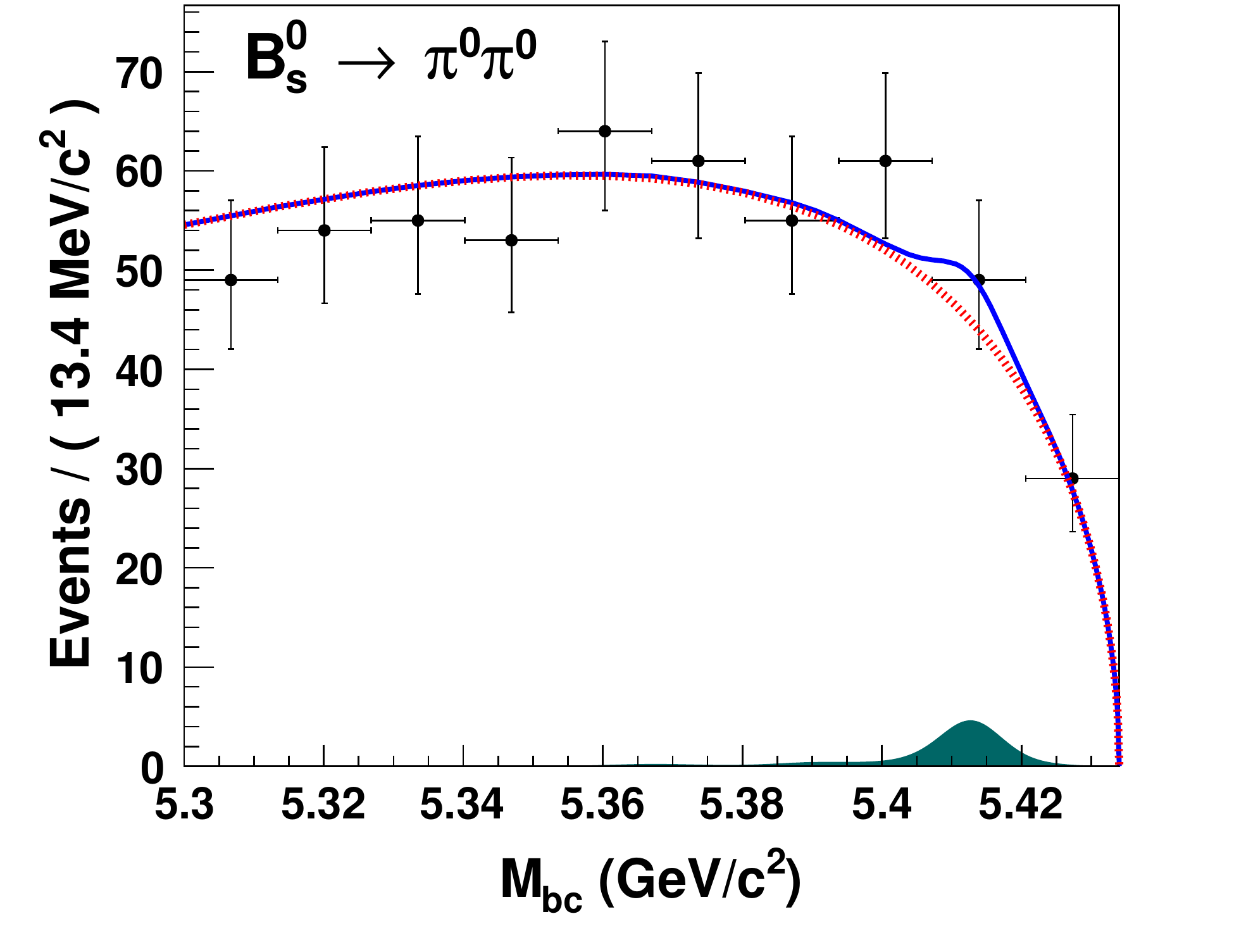}%
\includegraphics[height=4.7cm]{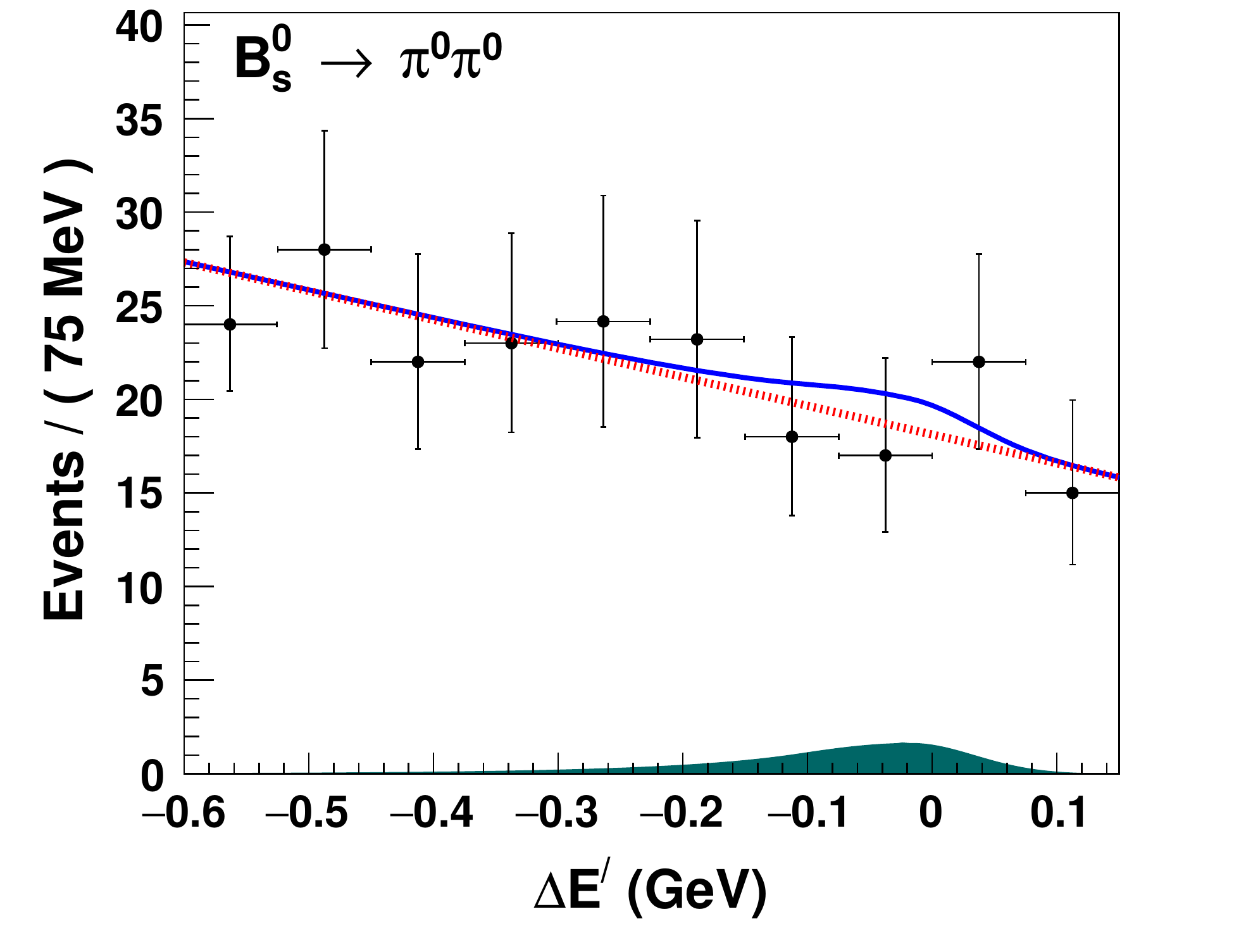}%
\includegraphics[height=4.7cm]{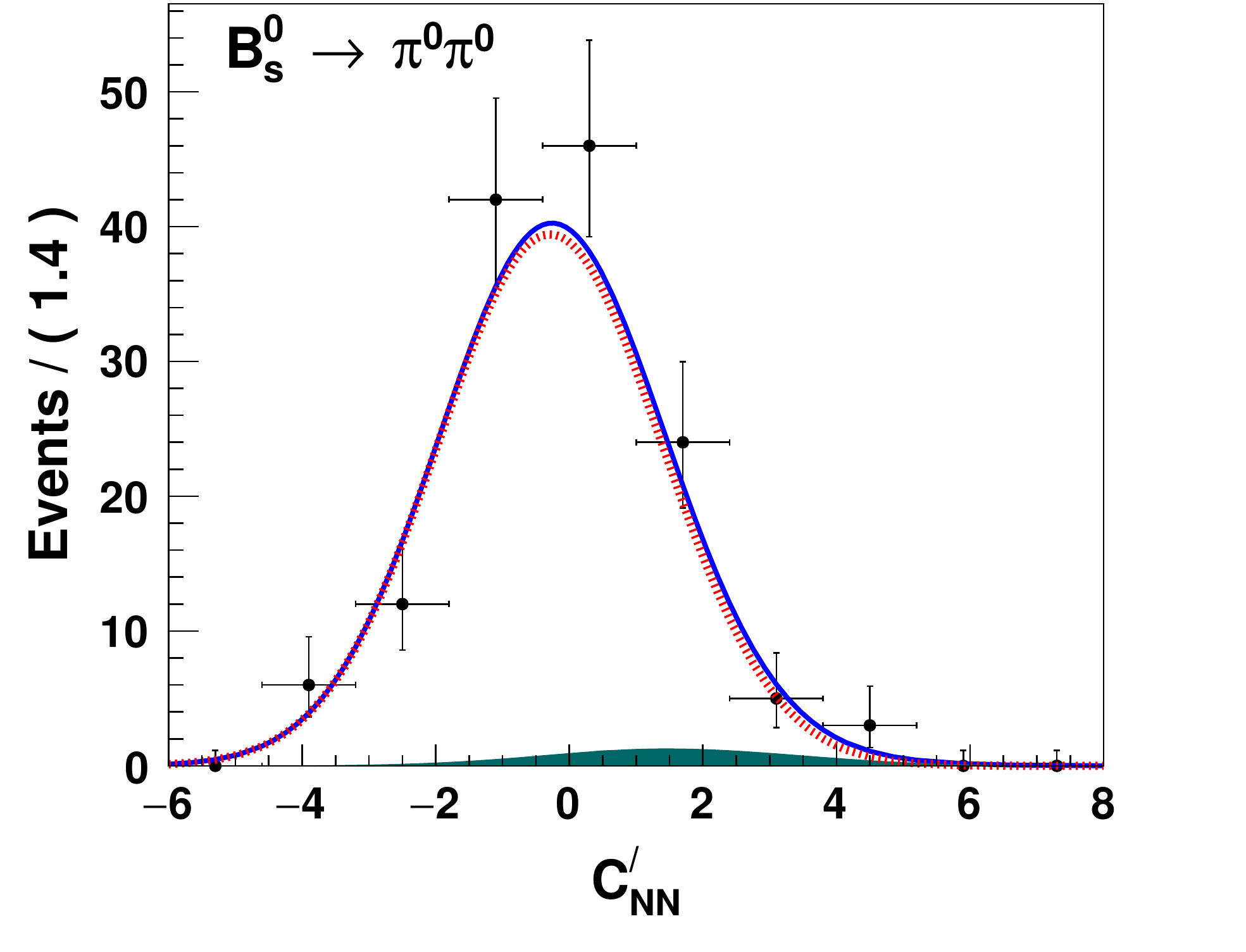} 
\caption{Signal enhanced projections of $M_{\rm bc}$ (left), $\Delta E^{\prime}$ (middle), and $\mathcal{C}^{\prime}_{\mathrm{NN}}$ (right) for the analysis, $B^0_s \rightarrow \pi^0\pi^0$. Each plot is generated by applying the signal region selection criteria on the two variables other than the plotted variable. The signal regions for the three variables are as follows, $5.395 \ \text{GeV}/c^2 < M_{\rm bc} < 5.434 \ \text{GeV}/c^2$ , $-0.310 \ \text{GeV} < \Delta E^{\prime} < 0.140 \ \text{GeV}$, and $-3.901 < \mathcal{C}^{\prime}_{\mathrm{NN}} < 7.451$. The dark-filled, red (dotted), and blue (solid) color distributions represent the signal, continuum background and total fit function, respectively. Points with error bars represent data. The peak in the $M_{\rm bc}$ distribution is due to the dominant $B^0_s$ production channel, $B^{*0}_s\bar{B}^{*0}_s$($87.0\%$). The other two production channels, $B^{*0}_s\bar{B}^0_s$($7.3\%$) and $B^0_s\bar{B}^0_s$($5.7\%$) are present, but suppressed in the plot.}
\label{data_fit}
\end{figure*}

To extract the signal yield, we perform a three-dimensional (3D) unbinned extended maximum likelihood (ML) fit to $M_{\rm bc}$, $\Delta E^{\prime}$, and $\mathcal{C}^{\prime}_{\mathrm{NN}}$. The likelihood function is defined as
\begin{linenomath}
\begin{equation}
\mathcal{L}_{\rm fit} = e^{-\sum\limits_{j} n_j} \mathbf{\prod_{\textit{i}}^{\textit{N}}} \left(\sum\limits_{j} n_j P_j ( (M_{\rm bc})^i, (\Delta E^{\prime})^i, {(\cal{C}'_{\mathrm{NN}}})^i)\right),
\end{equation}
\end{linenomath}
where $P_j ( M_{\rm bc}, \Delta E^{\prime}, {\cal{C}'_{\mathrm{NN}}})$ is the PDF of the signal or background component (specified by index $j$), $n_j$ is the yield of this component, $i$ represents the event index, and $\textit N$ is the total number of events in the sample. 

The linear correlation coefficients among  $M_{\rm bc}$, $\Delta E^{\prime}$, and $\mathcal{C}^{\prime}_{\mathrm{NN}}$  are found to be below $3\%$ in the signal region. Consequently, each of the 3D PDFs describing the signal and background contributions are assumed to factorize as $P_j \equiv P_j( M_{\rm bc})P_j (\Delta E^{\prime}) P_j(\cal{C}'_{\mathrm{NN}})$.
These factorized PDFs are modelled using large signal and background MC samples. The signal $M_{\rm bc}$ PDF consists of three PDFs corresponding to the three $B^0_s$ production channels. Each of them is again separately modelled from large MC samples. They are then combined according to their production fractions \cite{Sevda} to produce the final signal PDF for the $M_{\rm bc}$ variable. The PDF used for parametrizing $B^0_s\bar{B}^0_s$ is a sum of two Gaussian distributions with a common mean, while each of $B^0_s\bar{B}^{*0}_s\ \textrm{or}\ B^{*0}\bar{B}^0_s$, and $B^{*0}_s\bar{B}^{*0}_s$, are parametrized using a sum of a Gaussian function and an empirical PDF shape known as the Crystal Ball function \cite{CBall}. The signal $\Delta E^{\prime}$ variable, for all the three $B^0_s$ channels, is modelled using the Crystal Ball function, which is modified for this analysis to include the asymmetric nature of the distribution about the mean position. The output from the NN is parametrized using a Gaussian and an asymmetric (bifurcated) Gaussian PDF for the signal $\mathcal{C}^{\prime}_{\mathrm{NN}}$ variable. Unlike the signal PDF parameters for the $M_{\mathrm{bc}}$ variable, which is different for the three $B^0_s$ sources, $\Delta E^{\prime}$ and $\mathcal{C}^{\prime}_{\mathrm{NN}}$ variables take the same parameter values for the three $B^0_s$ production channels. The continuum background distribution of the $M_{\rm bc}$ variable is modelled through an empirically determined parametrized background shape referred to as the ARGUS function \cite{Argus}. The continuum background is parametrized using a first-order Chebychev polynomial and a sum of two Gaussian distributions for the $\Delta E^{\prime}$ and $\mathcal{C}^{\prime}_{\mathrm{NN}}$ variables, respectively. All the signal parameters and the background ARGUS endpoint are fixed to their best fit values obtained from $1\rm D$ fits to the MC simulated events. In contrast, all other background parameter values and the signal and background yields are floated. The PDFs used for modelling the signal and continuum background are listed in Table \ref{tab:table1}.

\begin{table}[htb]
\caption{\label{tab:table1} PDFs used to model the $M_{\rm bc}$, $\Delta E^{\prime}$, and $\mathcal{C}'_{\mathrm{NN}}$ distributions. The notations G, BG, CB, ACB,  CP, and A correspond to Gaussian, Bifurcated Gaussian, Crystal Ball, Asymmetric Crystal Ball, Chebyshev polynomial, and ARGUS functions, respectively.}
\begin{ruledtabular}
\begin{tabular}{ l l l l}
Fit component & $M_{\rm bc}$ & $\Delta E$ & $\cal{C}'_{\mathrm{NN}}$ \\
\hline
\small{
Signal} & G + G ($B^0_s\bar{B}^0_s$)& ACB & G + BG\\
        & G + CB ($B^0_s\bar{B}^{*0}_s$) & & \\
        & G + CB ($B^{*0}_s\bar{B}^{*0}_s$) & & \\
\hline
Continuum & A & CP & G + G \\
\end{tabular}
\end{ruledtabular}
\end{table}

To validate our analysis, we use the Belle data sample collected at the $\Upsilon (4\rm S)$ to reconstruct the decay $B^0_d \rightarrow \pi^0 \pi^0$ by applying similar event selection criteria. The results of the fit to $\Upsilon (4 \rm S)$ 
data are shown in Fig.~\ref{Bdpi0pi0}, where each fit projection is plotted after additional selection criteria are applied as described in the caption. We calculate the branching fraction, $\mathcal{B} (B^0_d \rightarrow \pi^0 \pi^0) = (1.18 \pm 0.21) \times 10^{-6}$ (where only the statistical uncertainty is shown), which is in good agreement with our previous result \cite{julius}.
 
The systematic uncertainties associated with the analysis are summarized in Table~\ref{tab:table2}. The systematic uncertainties due to the fit model are determined via ensemble investigations. To carry out an ensemble study, we generate and simulate $500,000$ signal MC events. We randomly select signal events from this sample for different expected signal yields in data. In addition, background MC events are randomly extracted from the background PDFs based on the expected number of background events in the data. This MC sample that now has statistics equivalent to the expected yields in data is amplified by repeating the above procedure a thousand times. We then perform  $3\rm D$ unbinned extended ML fits on these one thousand pseudo-experiments to obtain pull distributions for each of the expected signal yields in data. The average deviation of a constant function fit to the mean of the pull values from the no bias condition is recognized as a fit bias.

We observe a fit bias of $-3.3\%$ and assign it as the corresponding systematic uncertainty. The uncertainty due to fixing the parameter values of the PDFs is determined by varying the best fit parameter values within $\pm1\sigma$ of their statistical uncertainties and measuring the deviation of the signal yields in data. We find a fractional systematic uncertainty of $^{+3.5\%}_{-5.2\%}$ from this source. Apart from fixing the signal PDF parameters and the background PDF's ARGUS endpoint, we have also fixed the fractions of the $B^0_s$ production channels. We vary these fractions within $\pm1\sigma$ of their measured values \cite{Sevda} and repeat the fit. The observed relative variation $^{+5.2\%}_{-3.5\%}$ of the signal yield is assigned as the systematic uncertainty. The systematic uncertainty of the signal reconstruction efficiency is $0.4\%$ due to the finite number of signal MC events. The systematic uncertainty due to the efficiency of $\mathcal{C}^{\prime}_{\mathrm{NN}}$ requirement is estimated from the control sample using a parameter, $\mathcal{R}$. It is defined as the ratio between the efficiency of $\mathcal{C}^{\prime}_{\mathrm{NN}}$ in data and MC. We assign a corresponding systematic uncertainty of $\pm3\%$ due to the choice of the selection criteria on the NN output. 
 
The systematic uncertainty for the $\pi^0$ selection efficiency is determined to be $2.2\%$ per $\pi^0$ using the decay $\tau^- \rightarrow \pi^- \pi^0 \nu_{\tau}$. Since this uncertainty is completely correlated for the two $\pi^0$s, a total systematic uncertainty of $4.4\%$ is assigned. We assign a fractional systematic uncertainty of $0.03\%$ on the branching fraction of $\pi^0 \rightarrow \gamma \gamma$ \cite{pdg2022}. 
The systematic uncertainty due to the $b\bar{b}$ production cross-section at $\Upsilon(5\rm S)$ resonance, $\sigma_{b\bar{b}}$ is estimated to be $\pm4.7\%$ \cite{Sevda}. In addition, the systematic uncertainty due to the  three production charmless processes arising from $b\bar{b}$ events, $f_{\rm s}$ is assumed to be $\pm15.4\%$ \cite{pdg2022}. This uncertainty on $f_{\rm s}$ is the dominant systematic uncertainty associated with any $B^0_s$ measurement at Belle.

\begin{table}[htb]
\caption{\label{tab:table2} Summary of systematic uncertainties.
}
\begin{ruledtabular}
\begin{tabular}{l l}

Source & Value $(\%)$ \\
\hline
Fit bias & $-3.3$ \\
Fixed PDF parametrization & $^{+3.5}_{-5.2}$ \\
Fractions of $B^{*0}_s\bar{B}^{(*0)}_s$ & $^{+5.2}_{-3.5}$ \\
Reconstruction efficiency, $\epsilon_{rec}$ & $\pm0.4$ \\
$\mathcal{C}^{\prime}_{\mathrm{NN}}$ requirement & $\pm3.0$ \\
$\pi^0\rightarrow\gamma\gamma$ selection efficiency & $\pm4.4$ \\
$\mathcal{B}(\pi^0 \rightarrow \gamma \gamma)$  & $\pm0.03$ \\
$b\bar{b}$ cross-section, $\sigma_{b\bar{b}}$ & $\pm4.7$ \\
$f_s$ & $\pm 15.4$ \\
\hline
Total & $\mathbf{^{+18.1}_{-18.4}}$ \\ 
\end{tabular}
\end{ruledtabular}
\end{table}

The fit projections obtained from a 3D unbinned extended maximum likelihood fit in the signal regions are shown in Fig.~\ref{data_fit}. We obtain $5.7 \pm 5.8$ signal events and $989 \pm 32$ continuum background events in our fit to the data. The branching fraction is calculated using
\begin{linenomath}
 \begin{equation}\label{eq_bf}
\mathcal{B}(B^0_s \to \pi^0 \pi^0) =
\frac{N^{\rm sig}_{\rm yield}}{2 \times N_{B^0_{s} \bar{B}^0_{s}} \times  \epsilon^{\rm rec} \times  \mathcal{B}}
\end{equation}
\end{linenomath} where $N_{B^0_{s} \bar{B}^0_{s}}$ is the number of $B^0_{s} \bar{B}^0_s$ pairs; $\epsilon^{\rm rec}$ and $N^{\rm{sig}}_{\rm yield}$ are the signal selection efficiency obtained from MC simulation and the signal yield obtained from the fit, respectively; and $\mathcal{B}$ is the product of the two $\pi^0$-decay branching fractions  \cite{pdg2022}.

Incorporating the signal yield, $N^{\rm sig}_{\rm yield} = (5.7 \pm 5.8)$, number of $B^0_s\bar{B}^0_s$ pairs, $N_{B^0_s \bar{B}^0_s} = (8.30 \pm 1.34) \times 10^6$, the signal reconstruction efficiency, $\epsilon_{\rm rec} = (12.69 \pm 0.05)\%$, and branching fraction, $\mathcal{B}(\pi^0 \rightarrow  \gamma \gamma) = (98.82 \pm 0.03)\%$ in equation (\ref{eq_bf}), the branching fraction for  $B^0_s \rightarrow \pi^0 \pi^0$ and its product with $f_s$ are calculated to be
\begin{align*}
    \mathcal{B}(B^0_s \rightarrow \pi^0 \pi^0) &= (2.8 \pm 2.8 \pm 0.5) \times 10^{-6} \\
    f_{s} \times \mathcal{B}(B^0_s \rightarrow \pi^0 \pi^0) &= (0.6 \pm 0.6 \pm 0.1) \times 10^{-6}
\end{align*}
 The first uncertainty is statistical, and the second one is systematic.
 
Without significant signal yield, we calculate the UL on the branching fraction using a Bayesian approach. The UL on the branching fraction is estimated by integrating the likelihood function obtained from the maximum likelihood fit procedure from $0\%$ to $90\%$ of the area under the likelihood curve. The systematic uncertainties are incorporated by convolving the likelihood curve with a Gaussian distribution with a mean of zero and width equivalent to the total systematic uncertainty listed in Table~\ref{tab:table2}. The UL on the branching fraction, $\mathcal{B}(B^0_s \rightarrow \pi^0 \pi^0)$ at $90\%$ CL and the product of the branching fraction with $f_s$, $f_s \times \mathcal{B}(B^0_s \rightarrow \pi^0 \pi^0)$, is found to be
\begin{align*}
    \mathcal{B}(B^0_s \rightarrow \pi^0 \pi^0) &< 7.7 \times 10^{-6} \\
    f_s \times \mathcal{B}(B^0_s \rightarrow \pi^0 \pi^0) &< 1.5 \times 10^{-6}
\end{align*}
The total systematic uncertainties associated with $\mathcal{B}(B^0_s \rightarrow \pi^0 \pi^0)$ and $f_s \times \mathcal{B}(B^0_s \rightarrow \pi^0 \pi^0)$ are $^{+18.1\%}_{-18.4\%}$ and $^{+9.5\%}_{-10.0\%}$, respectively. The results are summarized in Table~\ref{tab:table3}.
 
 \begin{table}[htb]
\caption{\label{tab:table3}Summary of results on branching fractions and UL for $\mathcal{B}(B^0_{s} \rightarrow \pi^0 \pi^0)$ and $f_{s} \times \mathcal{B}(B^0_{s} \rightarrow \pi^0 \pi^0)$.}
\begin{ruledtabular}
    \begin{tabular}{l l}
        Quantity & Value \\
	\hline
	$\mathcal{B}(B^0_{s} \rightarrow \pi^0 \pi^0)$  &  $(2.8 \pm 2.8 \pm 0.5) \times 10^{-6}$ \\
		                                            & $< 7.7 \times 10^{-6}$ at $90\%$ CL      \\

	$f_{s} \times \mathcal{B}(B^0_{s} \rightarrow \pi^0 \pi^0)$                                   
		                                              & $(0.6 \pm 0.6 \pm 0.1) \times 10^{-6}$ \\
		                                              & $< 1.5 \times 10^{-6}$ at $90\%$ CL \\
		
    \end{tabular}
\end{ruledtabular}
\end{table}                                           
	
To summarize, we search for the decay $B^0_s \rightarrow \pi^0 \pi^0$ using the final Belle data sample available at $\Upsilon(5\rm S)$ resonance, which corresponds to an integrated luminosity of $121.4 \ \rm fb^{-1}$. We do not observe a significant signal yield, and thus set a $90\%$ CL upper limit on the $B^0_s \rightarrow \pi^0 \pi^0$ branching fraction of $7.7 \times 10^{-6}$. This is the most stringent UL estimated for this decay representing an order-of-magnitude improvement over the previous result \cite{L3} by the L3 experiment in 1995. 

\begin{center}
{\bf ACKNOWLEDGEMENTS}
\end{center}
This work, based on data collected using the Belle detector, which was
operated until June 2010, was supported by 
the Ministry of Education, Culture, Sports, Science, and
Technology (MEXT) of Japan, the Japan Society for the 
Promotion of Science (JSPS), and the Tau-Lepton Physics 
Research Center of Nagoya University; 
the Australian Research Council including grants
DP180102629, 
DP170102389, 
DP170102204, 
DE220100462, 
DP150103061, 
FT130100303; 
Austrian Federal Ministry of Education, Science and Research (FWF) and
FWF Austrian Science Fund No.~P~31361-N36;
the National Natural Science Foundation of China under Contracts
No.~11675166,  
No.~11705209;  
No.~11975076;  
No.~12135005;  
No.~12175041;  
No.~12161141008; 
Key Research Program of Frontier Sciences, Chinese Academy of Sciences (CAS), Grant No.~QYZDJ-SSW-SLH011; 
Project ZR2022JQ02 supported by Shandong Provincial Natural Science Foundation;
the Ministry of Education, Youth and Sports of the Czech
Republic under Contract No.~LTT17020;
the Czech Science Foundation Grant No. 22-18469S;
Horizon 2020 ERC Advanced Grant No.~884719 and ERC Starting Grant No.~947006 ``InterLeptons'' (European Union);
the Carl Zeiss Foundation, the Deutsche Forschungsgemeinschaft, the
Excellence Cluster Universe, and the VolkswagenStiftung;
the Department of Atomic Energy (Project Identification No. RTI 4002) and the Department of Science and Technology of India; 
the Istituto Nazionale di Fisica Nucleare of Italy; 
National Research Foundation (NRF) of Korea Grant
Nos.~2016R1\-D1A1B\-02012900, 2018R1\-A2B\-3003643,
2018R1\-A6A1A\-06024970, RS\-2022\-00197659,
2019R1\-I1A3A\-01058933, 2021R1\-A6A1A\-03043957,
2021R1\-F1A\-1060423, 2021R1\-F1A\-1064008, 2022R1\-A2C\-1003993;
Radiation Science Research Institute, Foreign Large-size Research Facility Application Supporting project, the Global Science Experimental Data Hub Center of the Korea Institute of Science and Technology Information and KREONET/GLORIAD;
the Polish Ministry of Science and Higher Education and 
the National Science Center;
the Ministry of Science and Higher Education of the Russian Federation, Agreement 14.W03.31.0026, 
and the HSE University Basic Research Program, Moscow; 
University of Tabuk research grants
S-1440-0321, S-0256-1438, and S-0280-1439 (Saudi Arabia);
the Slovenian Research Agency Grant Nos. J1-9124 and P1-0135;
Ikerbasque, Basque Foundation for Science, Spain;
the Swiss National Science Foundation; 
the Ministry of Education and the Ministry of Science and Technology of Taiwan;
and the United States Department of Energy and the National Science Foundation.
These acknowledgements are not to be interpreted as an endorsement of any
statement made by any of our institutes, funding agencies, governments, or
their representatives.
We thank the KEKB group for the excellent operation of the
accelerator; the KEK cryogenics group for the efficient
operation of the solenoid; and the KEK computer group and the Pacific Northwest National
Laboratory (PNNL) Environmental Molecular Sciences Laboratory (EMSL)
computing group for strong computing support; and the National
Institute of Informatics, and Science Information NETwork 6 (SINET6) for
valuable network support.


\begin{thebibliography}{99}
\bibitem{gronau}
M.~Gronau, D.~London, and J.~L.~Rosner, \href{https://doi.org/10.1103/PhysRevD.87.036008}{Phys. Rev. D {\bf 87}, 036008 (2013)}.

\bibitem{LHCb_PA}
R. Aaij {\it et al.} (LHCb Collaboration), \href{http://dx.doi.org/10.1103/PhysRevLett.118.081801}{Phys. Rev. Lett. {\bf 118}, 081801 (2017)}.

\bibitem{fit_QCDF}
Q.~Chang, J.~Sun, Y.~Yang, and X.~Li, \href{https://doi.org/10.1016/j.physletb.2014.11.027}{Phys. Lett. B {\bf 740}, 56 (2015)}. 

\bibitem{b2pp}
H.~Cheng, C.~Chiang, and A.~Kuo, \href{http://dx.doi.org/10.1103/PhysRevD.91.014011}{Phys.\ Rev.\ D {\bf 91}, 014011 (2015)}.

\bibitem{pqcd} 
A.~Ali {\it et al.}, \href{http://dx.doi.org/10.1103/PhysRevD.76.074018}{Phys.\ Rev.\ D {\bf 76}, 074018 (2007)}.

\bibitem{qcdf} 
M.~Beneke and M.~Neubert, \href{https://doi.org/10.1016/j.nuclphysb.2003.09.026}{Nucl.\ Phys.\ B {\bf 675} (2003) 333-415}.

\bibitem{L3}
M. Acciarri {\it et al.} (L3 Collaboration), \href{https://doi.org/10.1016/0370-2693(95)01042-O}{Phys.\ Lett.\ B {\bf 363}, 127 (1995)}.

\bibitem{charge_conjugate}
Charge conjugation is implied throughout this letter.

\bibitem{KEKB}
S. Kurokawa and E. Kikutani, \href{https://doi.org/10.1016/S0168-9002(02)01771-0}{Nucl.\ Instrum.\ Methods Phys. Res., Sect \ A {\bf 499}, 1 (2003)}, and other papers included in this Volume.
T. Abe {\em et al.}, \href{https://doi.org/10.1093/ptep/pts102}{Prog. Theor. Exp. Phys. 2013, 03A001 (2013) and references therein}.

\bibitem{detector1} 
A.~Abashian {\it et al.} (Belle Collaboration), \href{https://doi.org/10.1016/S0168-9002(01)02013-7}{
Nucl.\ Instrum.\ Methods Phys. Res., Sect \ A {\bf 479}, 117 (2002)}.

\bibitem{detector2} 
J.~Brodzicka {\it et al.} (Belle Collaboration), \href{https://doi.org/10.1093/ptep/pts072}{
PTEP {\bf 2012}, 04D001 (2012)}.

 
\bibitem{Sevda} 
S. Esen {\it et al.} (Belle Collaboration), \href{http://dx.doi.org/10.1103/PhysRevD.87.031101}{Phys.\ Rev.\ D {\bf 87}, 031101(R) (2013)}.

 \bibitem{pdg2022}
 R.~L.~Workman {\it et al.} (Particle Data Group), \href{https://doi.org/10.1093/ptep/ptac097}{Prog. Theor. Exp. Phys. \textbf{2022}, 083C01 (2022)}.

\bibitem{evtgen}
D. J. Lange, \href{https://doi.org/10.1016/S0168-9002(01)00089-4}{Nucl. Instrum. Methods Phys. Res., Sect. A {\bf 462}, 152 (2001)}.

\bibitem{geant3}
R. Brun {\it et al.} GEANT 3.21. Report No. CERN
DD/EE/84-1 (1984).

\bibitem{nn}
M. Feindt and U. Kerzel, \href{https://doi.org/10.1016/j.nima.2005.11.166}{Nucl. Instrum. Methods Phys. Res., Sect. A {\bf 559}, 190 (2006)}.
 

\bibitem{ksfw}
G. C. Fox and S. Wolfram, \href{https://doi.org/10.1103/PhysRevLett.41.1581}{Phys. Rev. Lett. {\bf 41}, 1581 (1978)}; S. H. Lee {\it et al.} (Belle Collaboration), \href{https://doi.org/10.1103/PhysRevLett.91.261801}{Phys. Rev.
Lett. {\bf 91}, 261801 (2003)}. 

\bibitem{bevan}
Ed.~A.~J.~Bevan, B. Golob, Th. Mannel, S. Prell, and
B. D. Yabsley, \href{https://doi.org/10.1140/epjc/s10052-014-3026-9}{Eur. Phys. J. C \textbf{74}, 3026 (2014)};
SLAC-PUB-15968; KEK Preprint 2014-3.  

\bibitem{punzi}
G.~Punzi, eConf C {\bf 030908} (2003), \href{https://doi.org/10.48550/arXiv.physics/0308063}{arXiv:physics/0308063 [physics.data-an]}.

\bibitem{CBall} 
T.~Skwarnicki, Ph.D. thesis, Institute for Nuclear Physics, Krakow, DESY Internal Report, DESY F31-86-02 (1986).

\bibitem{Argus} 
H. Albrecht {\it et al.} (ARGUS Collaboration), \href{https://doi.org/10.1016/0370-2693(90)91293-K}{Phys.\ Lett.\ B {\bf 241}, 278 (1990)}.

\bibitem{julius} T. Julius, M.E. Sevior, G.B. Mohanty, {\em et al.} (Belle Collaboration), \href{https://doi.org/10.1103/PhysRevD.96.032007}{Phys. Rev. D {\bf 96}, 032007 (2017)}.

\bibitem{Bayesian_declaration} As we use Bayesian method, this is ``credible interval'' but we use ``confidence level'' here following common convention.
\end{thebibliography}
\end{document}